\documentclass[a4paper, twocolumn]{article}
\usepackage{graphicx}
\usepackage{courier}
\usepackage{flushend,cuted}
\usepackage{amsmath}
\usepackage{url}
\usepackage{tikz}
\usepackage[american]{circuitikzgit}

\title{IREXF: Data Exfiltration from Air-gapped Networks by Infrared Remote Control Signals\thanks{Project Site: home.ustc.edu.cn/$\sim$zhou7905/IREXF}}
\author{Zheng Zhou, Weiming Zhang, Nenghai Yu\\
zhou7905@mail.ustc.edu.cn; \{zhangwm,ynh\}@ustc.edu.cn\\
University of Science and Technology of China\\
Key Laboratory of Electromagnetic Space Information\\ 
of Chinese Academy of Sciences} 
\begin{document}
\maketitle
\begin{strip}
\begin{abstract}
The technology on infrared remote control is widely applied in human daily life. It is also applied in the place with a top security level. Infrared remote control signal is regarded as a simple, safe and clean resource that can help us control the electrical appliances nearby. In this paper, we build IREXF, a novel infrared optical covert channel from a well-protected air-gapped network via a malicious infrared module implanted previously into a keyboard. A malware preinstalled in the air-gapped PC receives the data from the malicious infrared module to study the infrared surroundings in the air-gapped network. Once a suitable appliance is found, infrared remote control commands will be sent in a proper time. With the development of technology on Internet of Things, more and more electrical appliances can access Internet. Those infrared command signals exfiltrating out of the air-gapped network can be received by an appliance without any malicious configuration. In our experiment, via a smart TV set-top box, the rate of the covert channel can be up to 2.62 bits per second without any further optimization. Finally, we give a list of countermeasures to detect and eliminate this kind of covert channels.
\end{abstract}

\begin{flushleft}
\end{flushleft}
\end{strip}
\raggedend
\section{Introduction}
The concept of \textit{Covert Channel} was given by Lampson in 1973 to refer to those channels that are not used for normal communication.\cite{Lampson:1973:NCP:362375.362389} He found that the shared resources could be abused by the processes in different privilege levels to circumvent the security mechanism. With the development of network technology, many kinds of cover channel were found in past twenty years. Zander et al surveyed the network covert channels in different kinds of networks protocols in 2007. \cite{Zander2007-4317620}

In order to protect against the threats of network covert channels, physical isolation is conducted in almost every top-secret organization to keep the networks with high level separated from the less secure and Internet. This type of isolation is known as \textit{air-gapped}. Nevertheless, an air-gapped network is still not strong enough to eliminate the leakage of sensitive data.

A lot of methods were proposed to breach the air-gapped networks in the last decade. There are four kinds of covert channel to bridge the air gap: 
\begin{itemize}
\item Electro-magnetic covert channels,
\item Acoustic covert channels,
\item Thermal covert channels, and
\item Optical covert channels.
\end{itemize}
\textbf{Electro-magnetic covert channels:} Kuhn and Anderson proposed firstly the method\cite{kuhn1998soft} to transmit information covertly using electromagnetic radiation in 1998. 
Guri et al introduced AirHopper\cite{guri2014airhopper}, a type of malware, leak data between a mobile phone and a computer nearby using FM radio module in 2014. 
Guri et al introduced a malware named GSMem\cite{guri2015gsmem}, which leak data via electromagnetic radiation generated by the bus of computer memory in 2015.
Guri et al proposed USBee\cite{guri2016usbee}, which can be used to leak data via electromagnetic radiation generated by the USB cable in 2016.
In 2016, Matyunin et al used the magnetic field sensor in mobile device to build a covert channel.\cite{Matyunin2016}

\textbf{Acoustic covert channels:} In 2013, Hanspach and Goetz used the acoustical devices: speakers and microphones of the notebook computer to build a covert channel\cite{hanspach2014covert}.
Malley et al\cite{Malley-o2014bridging} introduced a covert communication over inaudible sounds in 2014.
Lee et al\cite{lee2015various} uses a loud-speaker as an acoustical input device, and make a speaker-to-speaker covert channel in 2015. 
Guri et al introduced Fansmitter\cite{Guri-Fansmitter-2016arXiv160605915G} and DiskFiltration\cite{Guri2017DiskFiltration}, new methods to send acoustic signals without speakers in 2016.

\textbf{Thermal covert channels:} In 2015, Guri et al introduced BitWhisper\cite{guri2015bitwhisper}, to build a unique bidirectional thermal covert channel via the heat radiated with another adjacent PC. 
In 2017, Mirsky et al proposed HVACKer\cite{Mirsky2017}, to build a one-way thermal covert channel from an air conditioning system to an air-gapped network.

The thermal covert channels in the multi-cores CPU is researched as follows: 
Mast built a thermal covert channel in multi-cores\cite{masti2015thermal} with a transmit rate of 12.5bits per second in 2015.
Bartolini studied the capacity of a thermal covert channel in multi-cores\cite{bartolini2016capacity} in 2016.
Selber propose UnCovert3\cite{selber2017uncovert3}, a new thermal covert channel in multi-cores with a transmit rate of 20 bits per second in 2017. 

\textbf{Optical covert channels:} Loughry and Umphres studied the exfiltration via LED indicators\cite{Loughry:2002:ILO:545186.545189} in 2002. They divided LED indicators into three classes. Class I includes the unmodulated LEDs used to indicate some state of the device; Class II is refer to the time-modulated LEDs correlated with the activity level of the device; Class III involves the modulated LEDs that are strongly correlated with the content of data being processed. They found that TD LED indicators on almost every modem of those years belong to Class III. Even an LED indicator on a DES encryptor leaks plain data. They indicated that although the LEDs in Class II are not so dangerous as those in Class III, but they can be modulated to leak significant signal, and can be used to build covert channels.

Sepetnitsky proposed a covert channel prototype \cite{Sepetnitsky-2014-6975588} of leaking data to the camera in a smart phone via the monitor's power status LED indicator in 2014.
Shamir presented a cover channel to breach an air-gapped network \cite{shamir2014light} by a light-based printer in 2014.
Lopes and Aranha proposed a malicious device\cite{lopes2017platform} to leak data via its flickering infrared LEDs.
In 2016, Guri introduced VisiSploit\cite{Guri2016An}, a prototype to leak data via an invisible QR-code in LCD screen.
Guri presented LED-it-GO\cite{Guri2017LED}, to leak data via hard drive LED indicators in 2017.
Guri also proposed xLED\cite{guri2017xled}, to leak data via status LED indicators on the routers in 2017. Zhou introduced KLONC\cite{Zhou-KLONC-2017arXiv171103235}, a prototype to leak sensitive information from an air-gapped network to a network surveillance camera via the status LED indicators on the keyboard in 2017.

There is a kind of optical covert channel to leak data via infrared signal that is invisible to human view. In 2016, Lopes and Aranha presented a malicious flash disk to leak data via infrared signal with a transmit rate of 15 bits per second.\cite{lopes2017platform}
Guri introduced a prototype to realize both exfiltration and infiltration of the data via infrared signal by circumventing an safety surveillance system.\cite{Guri-aIR-Jumper-2017arXiv170905742}

Lopes' prototype\cite{lopes2017platform} still needs a malicious internal person who holds a receiver to obtain the infrared signal. It is very difficult to be realized in the place with a high secret level.

\begin{figure}
\includegraphics[width=0.5\textwidth]{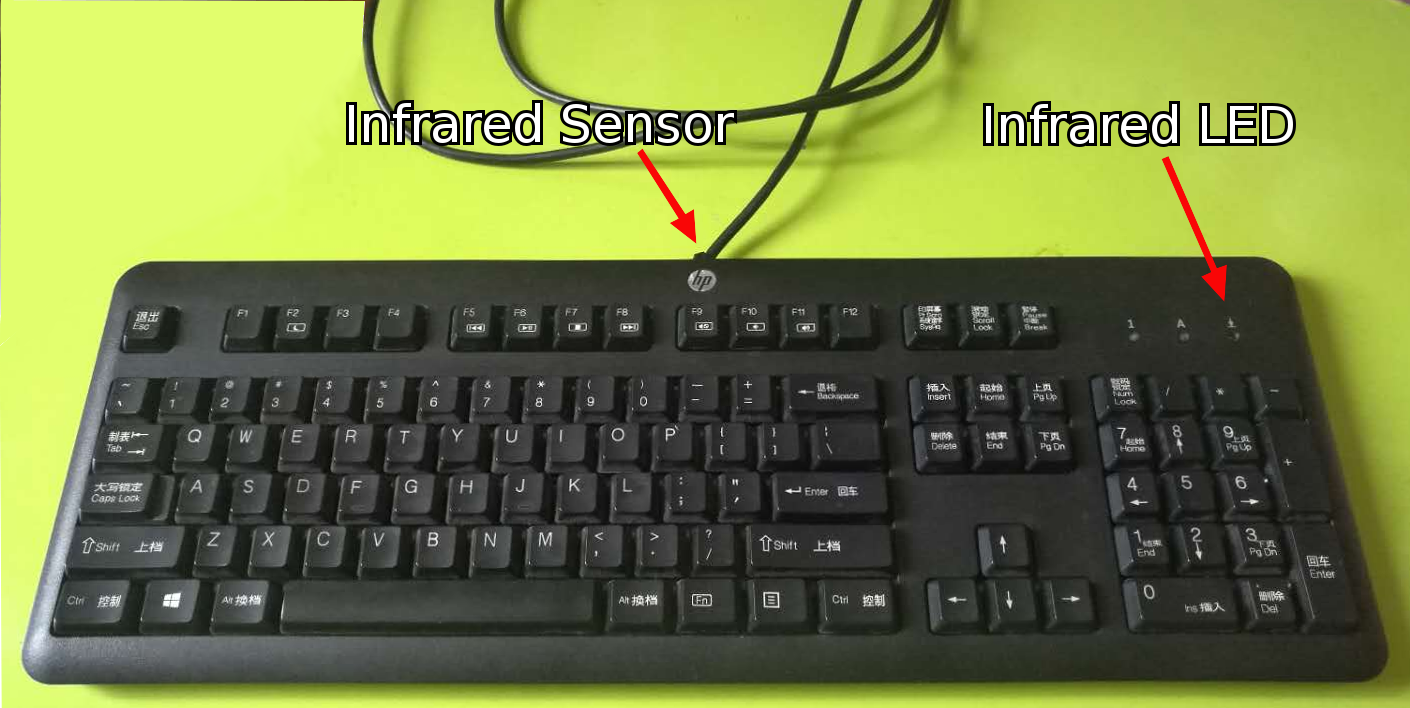} 
\caption{HP Keyboard Embedded with Malicious Hardware}
\label{keyboard}
\end{figure}

In our prototype, IREXF(the brief of ``infrared exfiltration''), a novel infrared optical covert channel is established from a well-pretected air-gapped network via a malicious infrared module implanted previously into a keyboard shown in Figure \ref{keyboard}. A malware preinstalled in the air-gapped PC receives the infrared signal from the malicious infrared module to study the infrared surroundings in the air-gapped network. After it finds a suitable appliance to control, it can select a proper time to send an infrared command. Then the infrared optical signal exfiltrates out of the air-gapped network. According the type of appliances, the sensitive data can be modulated into the behavior control infrared signals to the appliance.

Because the infrared signal is used for its original purpose. It is very difficult to distinguish between the real ones created from a remote controller by human being and the fake ones modulated from an infrared LED by the malicious hardware. It leads a lower probability to be checked out.

We find that in the background of IoT(Internet of Things), it is not necessary to control both sides of the covert channel as we do traditionally. As long as the operation method of a smart device is known, the device would be exploited maliciously rather than be hacked directly. 

The contributions of our research are as follows:
\begin{enumerate}
\item A set of malicious hardware is designed to fulfill the attack to an air-gapped network. It is available for a supply chain attack;
\item A prototype of data exfiltration from an air-gapped network via infrared remote control signals is proposed.
\end{enumerate}

The rest of the paper is organized as follows: Background technology is given in Section \ref{BackgroundTechnology}. A prototype, IREXF, is proposed in Section \ref{AttackModel}. Section \ref{ResultsandDiscussion} presents results and a discussion. Countermeasures are given in Section \ref{Countermeasures}, and we draw our conclusions in Section \ref{Conclusions}.

\section{Background Technology}\label{BackgroundTechnology}
\subsection{Infrared Transmition}\label{Infrared}
Infrared was discovered in 1800 by William Herschel. Infrared radiation is used in industrial, scientific, and medical applications, especially in short-ranged wireless communication. Infrared is the most common way for remote controls to command appliances.\cite{wikiInfrared} There are over ten infrared transition protocols: ITT, NEC, Nokia NRC, Sharp, Philips RC-5, Philips RC-6, Philips RECS-80 and Sony SIRC etc. The protocols are used to prevent surrounding infrared from interfering with the remote control signals.

Carrier wave is the key to distinguish the right signal to receive. For example, the infrared sensor HS0038B made by Vishay Telefunken can receive the signals with carrier frequency range of 35KHz to 41KHz with peak detection at 38KHz. Meanwhile, an HS0038B can receive all protocols. 

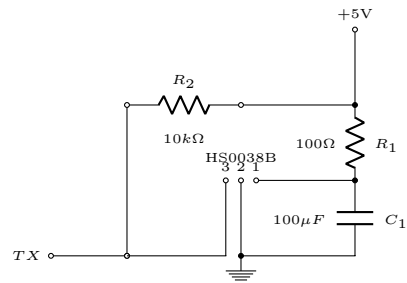
\begin{figure}[h!]
  \begin{center}
    \begin{circuitikz}
      \ctikzset{bipoles/length=.8cm}
      \draw (0,0) node[anchor=east]{\tiny{$TX$}} [short, o-] to (1,0) node[circ]{} 
to (1,2) to [R, l={\tiny{$R_2$}}, a={\tiny{$10k\Omega$}}](2.5,2)
to (4,2) node[circ]{}  [short, -o]to (4,3)  node[anchor=south]{\tiny{+5V}} ;
\draw (4,2) to [R, l={\tiny{$R_1$}}, a={\tiny{$100\Omega$}}](4,1)node[circ]{}
to [C, l={\tiny{$C_1$}}, a={\tiny{$100\mu F$}}](4,0)
to (2.5,0)node[circ]{}node[ground](gd){} [short, -o]to (2.5,1) node[anchor=south]{\tiny{2}};
\draw(4,1) [short, -o]to (2.7,1)node[anchor=south]{\tiny{1}};
\draw(1,0) to (2.3,0) [short, -o]to (2.3,1)node[anchor=south]{\tiny{3}};
\node at (2.5,1.3) {{\tiny{HS0038B}}};
    \end{circuitikz}
    \caption{Circuit to Receive Infrared Signal by HS0038B}
    \label{CircuitHS0038B}
  \end{center}
\end{figure}

The circuit for HS0038B is shown in Figure \ref{CircuitHS0038B}. The resistor $R_1$ is a current-limiting resistor, and resistor $R_2$ is
 a pull-high resistor to keep a high level on the node $TX$ while no infrared signal is received by HS0038B. The capacitor $C_1$ is a filter capacitor to maintain a constant voltage for HS0038B. A low level signal is output on the node $TX$ while any infrared signal is received.

Meanwhile, the infrared diodes can be used in an amplifier circuit.

\begin{figure}[h!]
  \begin{center}
    \begin{circuitikz}
    \ctikzset{bipoles/length=.8cm}
    \draw (1,2) node[npn](npn) at  (1,2) {\tiny{$S8050$}};
\draw (npn.B) to [R, l={\tiny{$R_1$}}, a={\tiny{$200\Omega$}}](-0.5,2)[short, -o] 
to (-1,2)node[anchor=east]{\tiny{$RX$}};
\draw (npn.E) to (1,1) node[ground](gd){};
\draw (3,1)node[anchor=north]{\tiny{+5V}} [short, o-]
to [empty diode, a={\tiny{$IN5819$}}] (3,2) 
to [full led, a={\tiny{$F3MM$}}] (3,3)
to [R, l={\tiny{$R_2$}}, a={\tiny{$20\Omega$}}] (1,3) 
to (npn.C);
    \end{circuitikz}
    \caption{Amplifier Circuit to Send Infrared Signal}
    \label{CircuitF3MM}
  \end{center}
\end{figure}
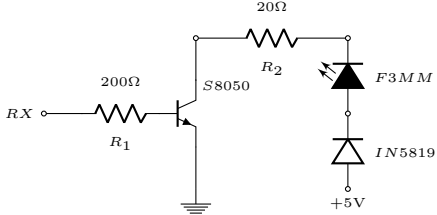  

In Figure \ref{CircuitF3MM}, the infrared light-emitting diode(LED) is labeled as $F3MM$. And $IN5819$ is a normal diode to protect the triode $S8050$ from an inverse voltage by any mistake. Both resistor $R_1$ and resistor $R_2$ are current-limiting resistors. the infrared diode $F3MM$ will send infrared radiation while a high level signal is kept on the node $RX$.
\subsection{Raspberry Pi}
The Raspberry Pi is a series of small single-board computers developed in the United Kingdom by the Raspberry Pi Foundation to promote the teaching of basic computer science in schools and in developing countries.\cite{wikiRaspberryPi} It serves as a controller for the infrared sensor and transmitter very well. And it is very easy to exchange data via its GPIO(General-Purpose Input/Output) bus with both sensors and the air-gapped PC.

The Raspberry Pi boots its OS since the power supply is available. Then it receives signals from the infrared sensor via GPIO continuously. 
\subsection{USB Adapter}
A USB adapter is a type of protocol converter which is used for converting USB data signals to and from other communications standards. Most commonly the USB data signals are converted to either RS232, RS485, RS422, or TTL-level UART serial data.
\cite{wikiUSBadapter} 

In our prototype, we use a USB-TTL adapter to simulate a serial port on the air-gapped PC. Anyone linking the port with correct parameters can receive or send data with TTL-level signals.

\section{Attack Model}\label{AttackModel}
\subsection{Scenario}
\begin{figure}
\includegraphics[width=0.5\textwidth]{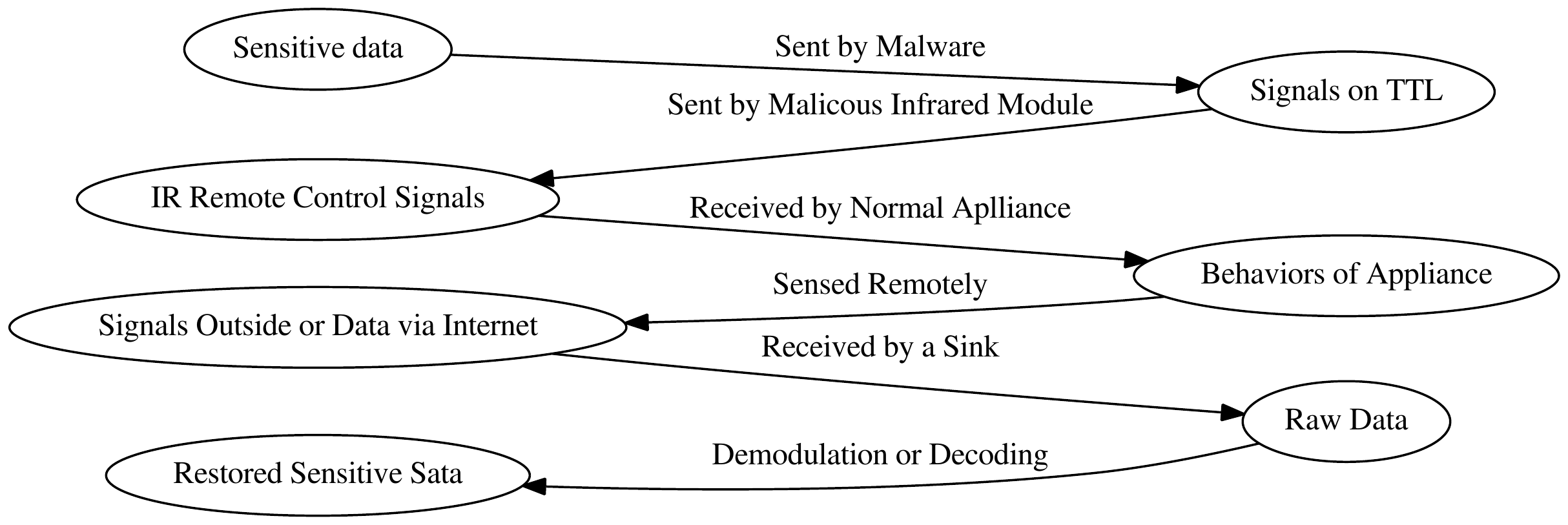} 
\caption{Flow diagram of IREXF}
\label{low_diagram_of_IREXF}
\end{figure}
An attack flow diagram is shown in Figure \ref{low_diagram_of_IREXF}, a malicious infrared module is implanted into a keyboard before the PC is given to its user(that is Supply Chain Attack), or when the PC is maintained by an attacker. Meanwhile, a malware is also preinstalled on the PC. The malware runs in the background while the PC is working. It seeks the sensitive data on the PC and copy them with a strong encryption. On the other hand, the malware receives data from the malicious infrared module to study the infrared surroundings in the air-gapped network. By checking up a signal table, it can judge the type and the brand of the electrical appliances, and find a suitable appliance to control. Then, at a proper time, it sends infrared remote control commands to the appliance and the infrared optical signals exfiltrate out of the air-gapped network. 

Because of the variety of the electrical appliances, the modulation forms of the signal are different. When the appliance is a traditional air-condition without accessing Internet, the signal can be modulated into the behaviors of sending a serious of infrared remote control commands. Then a receiver can be set out of the building near the air conditioner external unit; If the appliance is a smart air-condition that can be controlled via Internet, the signal can be modulated into some certain configuration value that can be easily read by an attacker who access the air-condition by an APP; If the appliance is a smart TV set, the signal can be modulated into a serious of infrared remote control commands to visit a malicious website with a long URL that involves the sensitive data.

In our prototype, IREXF, a smart TV box serves as a receiver. It is very easy to see this scenario in a control room with top secret: flat-panel TV sets are hung on the wall in front of some air-gapped PCs that are handled by the operator on duty.

\subsection{Transmitter}
\subsubsection{Malicious Infrared Module}
The malicious infrared module to send the infrared commands is made up of follows:

\begin{enumerate}
\item a USB-TTL adapter, which can simulate a serial port on the air-gapped PC;
\item a Raspberry Pi, which can 
\begin{itemize}
\item detect the surrounding infrared signals, and 
\item record the signals and transmit them to the air-gapped PC via TTL, and
\item send the infrared signals to infrared LED via GPIO.
\end{itemize}
\end{enumerate}

The malicious infrared module is link to the USB port on the PC in a hidden way. In our prototype, the module is embedded in a USB keyboard. Hence a USB hub is also needed to give another port.
\subsubsection{Malware on PC}
A malware is also required to be installed on the air-gapped PC. It can:
\begin{itemize}
\item fetch the sensitive information on the host
\item study the infrared signal from the malicious infrared module to make the infrared signal surrounding clear. Then it knows the type and the brand of the electrical appliances nearby.
\item give the malicious infrared module commands to send the infrared signal to a certain appliance. The sensitive data is sent by modulation in same time.
\end{itemize}
\subsubsection{Study on Infrared Signals}
In order to be compatible with more infrared transition protocols, our prototype is designed to record and replay a signal's waveforms directly rather than analysis its codes. The infrared signals are sniffed and compared with known waveforms to judge their protocols and the brands of the appliances. The results of judgments are restored in a database that can be queried by the malware in the air-gapped PC.

\subsection{Receiver}
A lot of appliances can be the receivers with the development of IoT. A receiver must has the following characters:

\begin{itemize}
\item can be controlled by infrared signals, and
\item can access Internet, or can be related with another appliance outdoor.
\end{itemize}

Any appliance meeting the two conditions above can be used as a receiver. After an investigation, we find that the smart TV sets, the smart air-conditions and the traditional air-conditions are most available in a high frequency. So, in our prototype, IREXF, a smart TV set-top box serves as the receiver.

\textbf{Smart TV set or TV with STB:} Android is the most popular mobile OS nowadays. More and more TV sets ship Android to enrich the functions and give a more wonderful experience to users. As a needful function, the network browser can access the websites on Internet. Naturally, the browser can be control by an infrared remote controller.

\textbf{Smart Air-condition:} Smart air-conditions become the mainstream, since a consumer can turn his air-condition on via Internet before he come back home. The temperature value and running mode etc. can be accessed and controlled with the APP on his mobile phone. Without question, a smart air-condition can also be controlled by an infrared remote controller.

\textbf{Traditional Air-condition:} Not only smart electrical appliances, the traditional ones can also be used to build a covert channel, as long as they have any relation with another appliance outside the building. Traditional air-conditions just meet this condition.
\section{Results and Discussion}\label{ResultsandDiscussion}
\subsection{Experiment Setting}
In our prototype named IREXF, we build a covert channel from an air-gapped PC to Internet or a lower security level network by infrared remote control signals that are modulated by a malicious hardware embedded previously with a supply chain attack.
\subsubsection{Hardware configuration}
\begin{figure}[h!]
  \begin{center}
    \begin{circuitikz}
    \ctikzset{bipoles/length=.8cm}
    \draw (1.5, -0.6) node[npn](npn) at  (1.5, -0.6) {\tiny{$S8050$}};
\draw (npn.B) to [R, l={\tiny{$R_1$}}, a={\tiny{$200\Omega$}},*-*](0, -0.6)[short, -o] to (-0.5, -0.6)node[anchor=east]{\tiny{RX}};
\draw (4,0)node[anchor=west]{\tiny{TX}} [short, o-]
to [empty diode, a={\tiny{$IN5819$}}] (3,0) 
to [R, l={\tiny{$R_2$}}, a={\tiny{$20\Omega$}},*-*] (2,0) [short, *-*]
to (npn.C);
\draw (-0.5,-1.5) node[anchor=east]{\tiny{GND}} [short, o-*] 
to (2,-1.5)node[circ]{}[short, *-o]
to (4,-1.5)node[anchor=west]{\tiny{GND}};
\draw (-0.5,-3.1) node[anchor=east]{\tiny{+5V}} [short, o-*] 
to (2,-3.1)node[circ]{}[short, *-o]
to  (4,-3.1)node[anchor=west]{\tiny{+5V}};
\draw (-0.5,-4) node[anchor=east]{\tiny{TX}} [short, o-*] 
to (2,-4)node[circ]{}[short, *-o]
to  (4,-4)node[anchor=west]{\tiny{RX}};
\draw (2,-1.5)[short,-*] to (npn.E);
\draw (2,-4) to [R, l={\tiny{$R_3$}}, a={\tiny{$10k\Omega$}}]
(2,-3.1) 
to [R, l={\tiny{$R_4$}}, a={\tiny{$100\Omega$}}]
(2, -2.25)node[circ]{} 
to [C, l={\tiny{$C_1$}}, a={\tiny{$100\mu F$}}](2,-1.5);
\draw (2, -2.25)[short, *-o]  to (4,-2.25)node[anchor=west]{\tiny{VCC}} ;
    \end{circuitikz}
    \caption{Circuit for Both F3MM and HS0038B}
    \label{CircuitforBothF3MMandHS0038B}
  \end{center}
\end{figure}
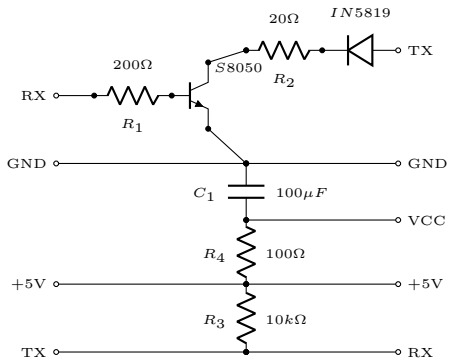

The malicious hardware is made up of follows:

\textbf{Infrared Receiver and Transmitter:} We use an infrared LED F3MM to replace the normal LED on the location of Scroll Lock, which is seldom used now. An infrared sensor HS0038B serving as an infrared signal sniffer can hides itself behind the translucent plastic panel on a keyboard or anywhere can receive optical signals easily. The circuits of them is redesigned into one shown in Figure \ref{CircuitforBothF3MMandHS0038B} to shorten its size into 1.9cm x 2.6cm.

\textbf{Raspberry Pi:} Among all type of Raspberry Pi, the Raspberry Pi Zero has the smallest size: 6.5cm x 3.0cm x 0.5cm. Hence it is available to be embedded in a normal-sized keyboard.

\textbf{Others:} A USB hub,  and a USB-TTL adapter are also needed.

\begin{figure}
\includegraphics[width=0.5\textwidth]{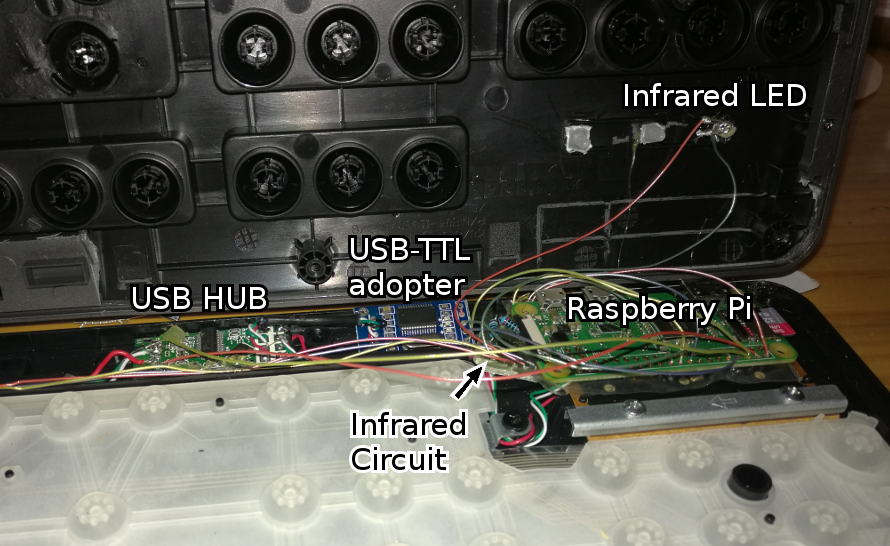} 
\caption{Inside the Keyboard with Malicious Hardware}
\label{keyboardinside}
\end{figure}

\textbf{Hardware Host Object:} An HP keyboard with model KU-1156 shown in Figure \ref{keyboardinside} is used as a host object to accommodate all the malicious hardware:
\begin{itemize}
\item HS0038B, F3MM, and their circuit board;
\item Raspberry Pi Zero and a USB-TTL adapter;
\item A USB hub.
\end{itemize}

\subsubsection{Sink of Covert Channel}

A Skyworth Network Set-Top Box Q+ 2th, which is one of the most popular TV set-top boxes in China, is used. 

A web site written in Django can be access by the TV box. It receives the covert data and restore them into an SQLITE database. 

\subsubsection{Experimental Environment}
In our prototype, we assume that there is a TV set with a network set-top box in the same room of an air-gapped PC. It is a common scenario in a control room: a row of computers are placed in the center of the room. Their screens, keyboards and mice are put on the desk, and their main boxes are put under the desk. Several panel TV sets are hung on the wall in front of the operators. Some of the panels are used to display the key information of the systems they control. One of the panels is used to play TV shows in real time. In the place with a high security level, any personal smart device is banned. So it is very necessary and reasonable to place a TV set to remove loneliness and disorientation of an operator on a long-timed duty.

\subsubsection{Conversation from Text to Remote Control Commands}
\begin{figure}
\begin{center}
\includegraphics[width=0.25\textwidth]{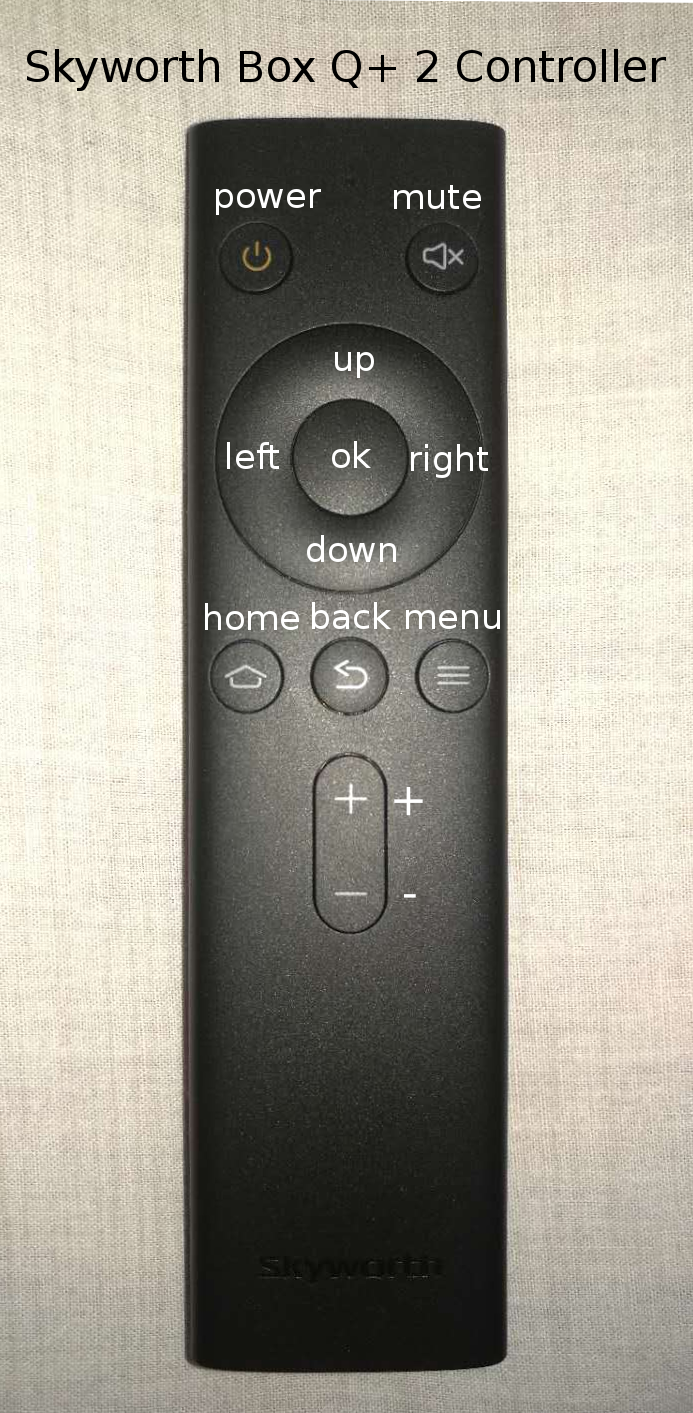} 
\caption{Controller of TV box}
\label{Controller_of_TV_box}
\end{center}
\end{figure}

The users of a TV box can not input text directly using a remote controller. The buttons on the controller of Skyworth TV box is shown in the Figure \ref{Controller_of_TV_box}. When a user want to input some words, a few buttons they can use such as `up', `down', `left', `right' and `ok'. An IME(Enput Method Editor) APP, such as Baidu TV IME will be set active to give a help to the user.

\begin{figure}
\begin{center}
\includegraphics[width=0.5\textwidth]{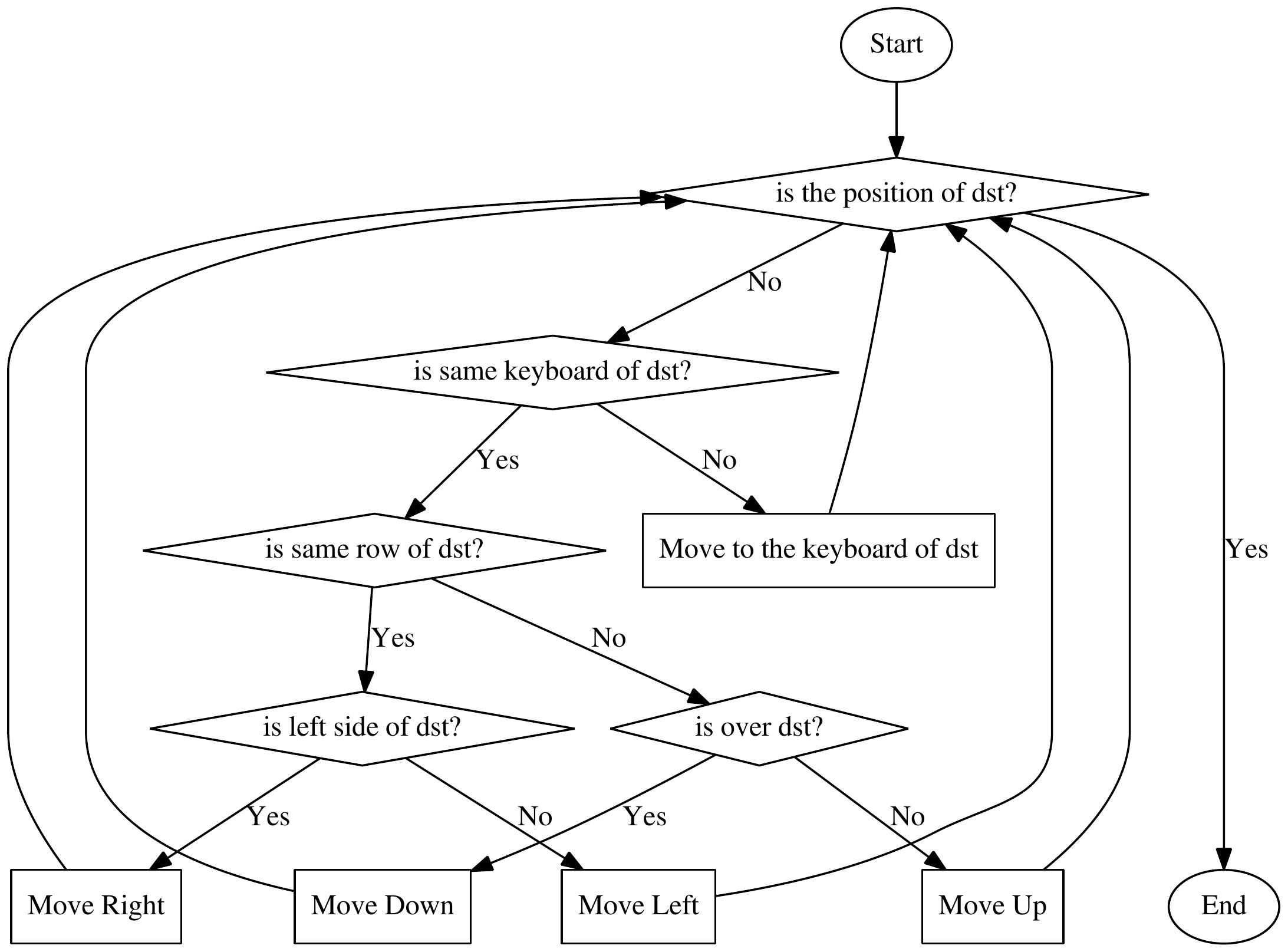} 
\caption{Flow diagram of Conversation Algorithm}
\label{FlowDiagramofConversationAlgorithm}
\end{center}
\end{figure}

So, we designed a conversation algorithm to convert the words(made up of numbers and letters) into a serials of remote control commands. In Baidu TV IME, there are four keyboards in English mode: lowercase letters, uppercase letters, numbers and symbols. We made four tables to record the locations of every byte. Then a path from a byte to another can be calculated out by considering the difference of their locations. The initial locations of the former three keyboards are always on `q', `Q' and `1'. The flow diagram of the conversation algorithm is shown in Figure \ref{FlowDiagramofConversationAlgorithm}.
\subsection{Results}
\subsubsection{Existence of Covert Channel}
In one of our experiments, we sent the string ``KeyboardisOKon20180104ThankyouUSTC'' with the length of 34 bytes. Firstly, a serials of commands would be sent to change the input source of the TV set and order the TV box to open the browser APP. It took 9 second to fulfill this procedure. Then we waited 10 seconds for the start of the APP. Secondly, we sent the commands to input the prefix in the URL textbox. The prefix is used to indicate the protocol and the site address. There were 173 commands sent in this procedure in 39 seconds. Thirdly, the payload of 34 bytes was sent. It cost 68 seconds. Finally, some commands are sent to close the browser APP and recovery the interface back to home and change the input source of the TV set back in 4 seconds.

We logged in the site accessed by the TV box. And the access record was found in our database:

\texttt{\small 15|2018-01-04 07:02:17.652151|\linebreak
KeyboardisOKon20180104ThankYouUSTC}

\subsubsection{Channel Rate}
\begin{figure}
\begin{center}
\includegraphics[width=0.5\textwidth]{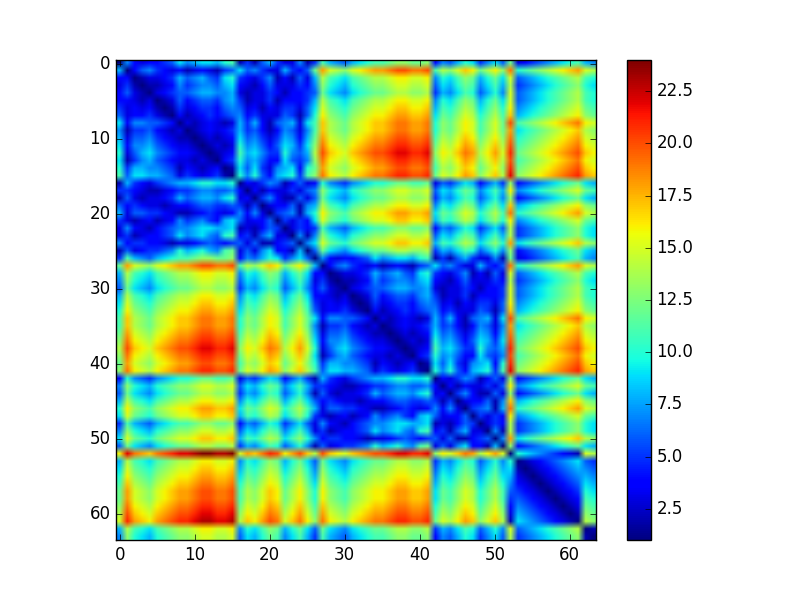} 
\caption{Command Numbers of Moving from one Byte to Another}
\label{CommandNumbers}
\end{center}
\end{figure}
As shown in Figure \ref{CommandNumbers}, the command numbers of moving from one byte to another byte according to the conversation algorithm designed ahead are calculated. The axises values from 0 to 63 stand for the bytes `A'$\sim$`Z', `a'$\sim$`z', `0'$\sim$`9', `-' and `\underline{\hspace{0.5em}}' which can be used to encode data with a modified Base64 for URL\cite{wikiBase64}. The numbers are various: the smallest one is 1, when the next byte is just the same one previous; the biggest one is 24, when they locates in different keyboard and they are both most far from the button to change the keyboard, such as the moving from `0' to 'L' etc.

The procedure to input a text can be consider as a Markov Process with 64 status. The transition probability matrix can be calculated by the statistic on the input texts. Then, the average of the numbers of moving can be obtained. The rate of this covert channel will be determined finally.

According to Information Theory, the rate can be defined as follow:

\[R=\frac{nH(X)}{t}\]

Where, $R$ is the rate, $H(X)=-\sum_{x_i \in X}P(x_i)\log P(x_i)$ is the information entropy of the source $X$, $x_i$ is a codeword of $X$, $n$ and $t$ mean that there are $n$ codewords of $X$ transmitted in the channel in $t$ seconds.

Especially, when all 64 bytes are used with the same probability($P(x_i)=\frac{1}{64}$), such as the data is encoded with Base64, the average of the numbers of moving is 10.147461. We know that one byte has 6($=-\log_2\frac{1}{64}$) bits information for Base64. Meanwhile, we measured that it cost 0.225433526 second on sending a remote control command successfully. Then, the rate is 2.622861484 bits per second.

\subsection{Discussion}
\subsubsection{Necessary Conditions}
The necessary conditions of the existence of this kind of covert channels are:
\begin{itemize}
\item There is at least one electrical appliance that can be controlled by infrared remote control commands.
\item There is at least one remote controller that is often used to send infrared signal which can be studied by the malicious hardware.
\item There is an adequate space in the keyboard to settle all parts of the malicious hardware.
\end{itemize}
\subsubsection{Error on Remote Control Commands}
When a string would be inputted, any error occurring on the transmission of remote control commands is a fatal one. Because the position of any followed byte depend on the position of the previous one. If there is not a pause to reset the position, a wrong position will impact all subsequent bytes.

In order to facilitate users to input text, Baidu TV IME introduce the function ``ring shift'' to make the moving of highlighted byte more quick. 

We find that there is no ``ring shift up'' when some bytes are already typed to select on the top row of IME interface. The highlighted area will stays on the fist alternative item, no mater how many `up's are clicked. Hence we can use this character to give a pause to the procedure of input. An array of commands: ``up, up, up, up, ok'' are wonderful to match all position in the keyboard. Then the new position is always the initial position(`q' or `Q') in current keyboard. A pause is used to reset the position to avoid continuous errors.
\subsubsection{Covertness}
There is an infrared receiver on the malicious hardware. It is used to study the infrared surrounding. But it also can be used to find an exceptional situation during the procedure of sending remote control commands. If a received infrared signal is different from the one we send, that means there is another party that sends the infrared command. It is almost certain that the party is a controller held by a people. So, it is not a good time to build a covert channel. A shell code must be executed to halt the sending procedure immediately. 
\subsubsection{For a Higher Rate}
In order to get a higher rate of the covert channel, we can keep on researching in the following aspects:
\begin{enumerate}
\item Try to shorten the average number of movings by using the ``roll shift'' functions in Baidu TV IME;
\item Try to shorten the number of commands by using the ``word auto-completion'' function in English mode of Baidu TV IME;
\item Try to increase the input speed with a shorter gap between commands;
\item Shorten the prefix of URL with a shorter domain name.
\end{enumerate}

\section{Countermeasures}\label{Countermeasures}
The countermeasures are threefold: design countermeasures, procedural countermeasures and technical countermeasures.

Design countermeasures can increase the difficulty of exploitation on the appliances receiving infrared commands. Banning web browser APPs from a TV Box is designed in some TV set-top boxes, such as Tmall MagicBox, a famous TV box brand in China. But the market occupancy of Tmall MagicBox has been falling down since the manufacturer banned all the third party APPs from their boxes.

Procedural countermeasures involve banning Internet from a smart air-condition and covering the LEDs unused. Both banning Internet and covering the LEDs are easy to applied, but they make users inconvenient.

Technical countermeasures involve infrared signals monitoring with hardware and redundant device detecting. The aim of the former is to sniff the infrared signals by a set of infrared receiver. The latter is used to detect the existence of malicious hardware. But it is difficult to find the covert channel out with both of them, since the covert channel is not in work all the time. Besides, it is also hard to distinguish the behaviors from a normal remote controller and the ones from a malicious hardware, while human behaviors are simulated exactly.

The list of all countermeasures is summarized in Table \ref{CountermeasuresList}.

\begin{table*}
\caption{Cost and Effect of Countermeasures}
\label{CountermeasuresList}
\begin{tabular}{l|c|c|c|l}
\hline 
Countermeasure & Type & Cost & Effect & shortcomings\\
\hline 
Banning web browser APPs from a TV Box & Design & Low & Good & Bad image to consumers \\
\hline
Banning Internet from a smart air-condition & Proc. & Low & Good & Inconvenience to user\\
Covering the LEDs unused & Proc. & Low & Poor & Inconvenience to user\\
\hline
Infrared signals monitoring with hardware & Tech. & High & Normal & Difficult to find\\
Redundant device detecting & Tech. & Low & Normal & Difficult to find\\
\hline 
\end{tabular}
\end{table*}

\section{Conclusions}\label{Conclusions}
A novel method on data exfiltration from an air-gapped network by infrared remote control signals was proposed in this paper. With this method, the normal appliance can be controlled remotely to help the leakage of sensitive information covertly.
An attack model, IREXF, was introduced to build a covert channel with a smart TV set-top box controlled by infrared remote control signals sent by a malicious hardware embedded in the keyboard. 
The detection of ambient appliances controlled by infrared signals was studied. 
In order to send more ASCII codes to the covert channel, the conversation from a text string to a set of remote control commands was designed.
The result of experiment shows that the rate of IREXF can be up to 2.62 bits per second.
The relative issues on this kind of covert channels are discussed. 
Finally, countermeasures were given by considering the key points on the flow of this kind of covert channel.

\Urlmuskip=0mu plus 1mu\relax
\bibliographystyle{plain}

\end{document}